\documentclass[twocolumn,aps,showpacs]{revtex4}

\usepackage{graphicx}
\usepackage{dcolumn}
\usepackage{amsmath}

\begin{document}

\title{Barkhausen Noise in a Relaxor Ferroelectric}
\author{Eugene V. Colla}
\author{Lambert K. Chao}
\author{M. B. Weissman}
\email{mbw@uiuc.edu}
\affiliation{Department of Physics, University of Illinois at
Urbana-Champaign, 1110 West Green Street, Urbana, IL 61801-3080
}

\date{\today}

\begin{abstract}
Barkhausen noise, including both periodic and aperiodic components, is
found in and near the relaxor regime of a familiar relaxor
ferroelectric, PbMg$_{1/3}$Nb$_{2/3}$O$_3$, driven by a periodic
electric field. The temperature dependences of both the amplitude and
spectral form show that the size of the coherent dipole moment changes
shrink as the relaxor regime is entered, contrary to expectations based
on some simple models.
\end{abstract}


\maketitle

Although relaxor ferroelectrics form locally ferroelectric polar
nanodomains, underlying chemical disorder somehow prevents the formation
of long-range ferroelectric or antiferroelectric order
\cite{Cross:reviews}. Instead, there is a crossover, characterized by
faster-than-Arrhenius temperature dependence of relaxation rates
(approximately of the Vogel-Fulcher form, e.g.\ \cite{Levstik:98}), to a
glassy ``relaxor'' regime. The key ingredients of a successful model may
include static random polar fields (e.g.\ \cite{Westphal:92-Colla:95}),
random anisotropy (e.g.\ \cite{Egami:99}), random interactions among the
polar nanodomains (e.g.\
\cite{Egami:99,Viehland:91,Blinc:99,Vugmeister:00}), and interactions
with soft TO phonons \cite{Gehring:00,Vikhnin:00FE,Vugmeister:00} and
slow charge-transfer modes \cite{Vikhnin:00FE}.

Nonequilibrium noise and mesoscopic noise provide model-sensitive probes
of glassy freezing \cite{Weissman:96,OBrien:PLZTBkh}, unlike macroscopic
linear response and the corresponding equilibrium fluctuations, which
show rather generic properties. The Barkhausen effect (see, e.g.,
\cite{Rudyak:70}) arises because the polarization changes unevenly as an
applied field changes, with the random part of the response only
statistically similar between nominally identical samples. For two
materials with similar average response functions, the Barkhausen noise
will be larger for the material whose polarization makes larger coherent
changes.

Although simple pictures of interacting polar nanodomains would predict
that the dynamically coherent units grow as the frozen regime is
entered, a previous attempt to find Barkhausen noise in the relaxor
regime found none, setting an upper limit on the cooperative changes of
the dipole moment \cite{OBrien:PLZTBkh}. Large discrete steps in polar
order have been found in the best-studied relaxor,
PbMg$_{1/3}$Nb$_{2/3}$O$_3$ (PMN), but only after the applied dc
electric field was large enough to drive the material into an ordinary
ferroelectric regime with macroscopic domains
\cite{Westphal:92-Colla:95}. Here we report Barkhausen noise in the
\emph{nonferroelectric} (paraelectric and relaxor) regimes of PMN, where
conventional Barkhausen noise from macroscopic domains is absent.

The simplest Barkhausen model \cite{Preisach:35} consists of a set of
polarization steps at fixed fields distributed with
uniform probability over the field sweep range. For periodically driven
fields, the noise consists of a periodically repeated random walk, so
$S(f)$, the Fourier power spectral density of the voltage fluctuations,
will consist only of harmonics of the driving frequency $f_D$, and its
spectral envelope will decay as $1/f^2$. If the step pattern changes
between cycles (e.g., due to thermal jitter in the step times) then
there will be aperiodic noise too. For fast-relaxing domains which
remain in quasi-equilibrium as the field changes, the polarization steps
follow the smooth change in Boltzmann factor vs. time (on a coarse time
scale), causing the $1/f^2$ envelope to cut off rapidly
($\textrm{csch}^2(\pi f/f_C)$) above a characteristic frequency $f_C$
\cite{OBrien:PLZTBkh}:
\begin{equation} \label{eq:1f2Cutoff}
f_C \equiv f_D \frac{p E_D}{k_B T} \quad .
\end{equation}
(Here, $E_D$ is the maximum amplitude of the sinusoidal driving field,
$p$ is the typical change in electric dipole moment in a single step,
and $k_B$ is Boltzmann's constant.) For domains with large barriers to
switching, the switches occur as single abrupt steps. However, thermal
jitter in the timing of otherwise reproducible steps also
reduces the periodic component above $f_C$, with a (numerically
computed) envelope very similar to that for the fast switchers. The
missing harmonic power should appear as an aperiodic component of
roughly Lorentzian shape and corner frequency comparable to $f_C$.

When this simple independent-step picture applies, the voltage variance
(the integral of $S(f)$) will grow linearly with $E_D$ (so long as $f_C
\gg f_D$) \cite{Rudyak:70}, being \cite{OBrien:PLZTBkh}
\begin{equation} \label{eq:TotBkhNoise}
\langle (\delta V)^2 \rangle \approx \frac{p E_D}{\varepsilon C_o} \quad .
\end{equation}
Here $C_o$ is the geometrical capacitance, and we assume the dimensionless
dielectric constant $\varepsilon \gg 1$.  When there is a range
of $p$'s, $p$ in Eq.~\ref{eq:TotBkhNoise} is replaced by $\langle
p^2 \rangle / \langle p \rangle$.

Generically, if the non-Arrhenius slowing down of relaxation rates
arises because interacting nanodomains form clusters which reorient
coherently but with barriers which increase with cluster size (e.g.\ as
in the random-field Ising model \cite{Huse:87}), then the dipole moments
of the coherent clusters (even if the constituents are randomly aligned)
would grow as $T$ was reduced, causing $\langle (\delta V)^2 \rangle$
and $f_C$ to rise. If the activation barrier heights for some clusters
become too large for them to respond at or near $f_D$ for a given $E_D$,
then the typical $p$ would saturate near the moment of the largest
responding clusters. The size of the largest responding cluster would
shrink slightly as $T$ is reduced further, but only as $T$ to a power
less than 1. We found neither continued growth nor approximate
saturation of the noise magnitude and frequency range, but instead
abrupt shrinkage.

Our sample consisted of two pieces, each about 0.75~mm thick, from a
single crystal of PMN, grown by the Chokhralsky method at the
Rostov-on-Don Institute of Physics. The sputtered gold contact pads on
the main (001) faces had about 1~mm$^2$ surface area each.

\begin{figure}[!tb]
\begin{center}
\includegraphics[width=0.4\textwidth,clip]{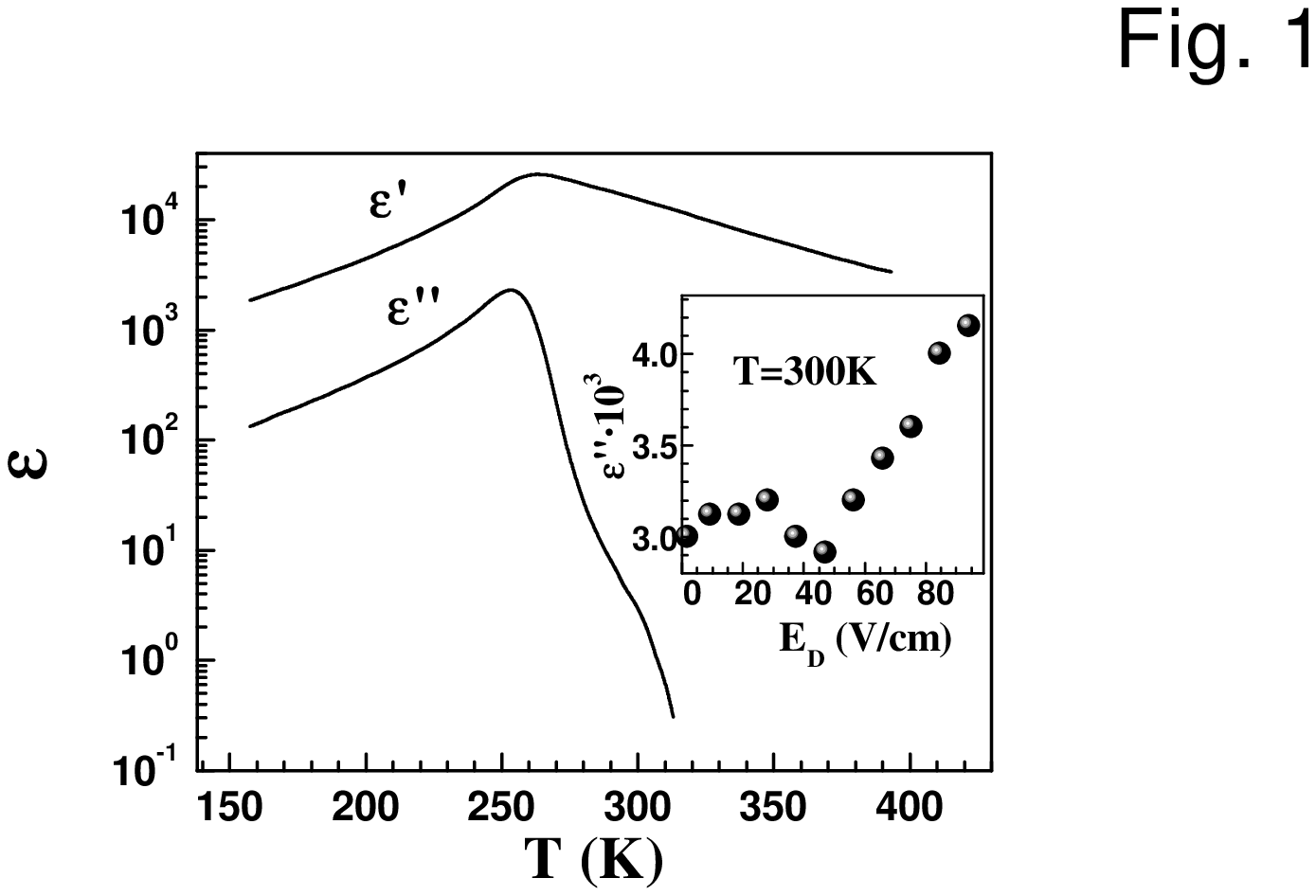}
\end{center}
\caption{
$\varepsilon '(T)$ and $\varepsilon '' (T)$ at 50~Hz, measured at
13~V/cm. The inset shows the nonlinear out-of-phase response, most
apparent at high-$T$ where the linear out-of-phase response is tiny.}
\label{fig1}
\end{figure}

Standard ac susceptibility measurements at 50~Hz showed the usual PMN
relaxor behavior, as shown in Fig.~\ref{fig1}. The response was
close enough to linear over the field range employed in the noise
experiments for the linear susceptibility to suffice for approximate
calculations. Previous experiments \cite{Colla:99} have shown that the
apparent Vogel-Fulcher freezing temperature of PMN ($\approx 220$~K)
determined by ac response is essentially unchanged over the field range
used in these experiments, although it drops sharply at higher fields.

\begin{figure}[!tb]
\begin{center}
\includegraphics[width=0.4\textwidth,clip]{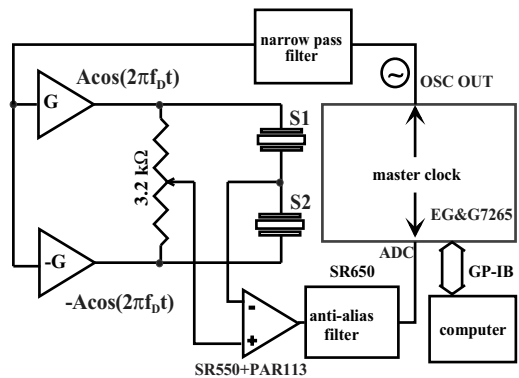}
\end{center}
\caption{The schematic diagram of the Barkhausen measurement circuit. A
trimming capacitor in the balance circuit has been omitted to avoid
clutter. S1 and S2 are the sample halves. $\pm$G is the gain of the
driving op-amps. Data were taken on the 14-bit converter in sets of
30,000 points with a 10~kHz sampling rate, giving 1/3~Hz frequency bins
in the Fourier spectrum. Usually 16 such spectra were averaged. The
anti-alias filter corner frequency was 5~kHz, although it should
have been 4~kHz.}
\label{fig2}
\end{figure}

For the noise measurements, the two halves of the sample were
incorporated in a balanced-bridge circuit, as shown in Fig.~\ref{fig2}.
This setup allowed us to apply an ac bias up to 10~volts on each half of
the sample without overloading the low-noise amplifiers, despite the
generation of harmonics by the sample nonlinearity, slow capacitance
changes due to systematic aging effects \cite{Colla(mbw):aging}, and imperfect
common-mode rejection. However, any residual systematic differences
in the nonlinearity of the two parts would generate harmonics. An experiment
in which one sample arm was replaced with a polystyrene capacitor indicated
significant systematic harmonics out to the fifth harmonic. We avoided
using harmonics lower than the tenth in the data analysis, since
systematic nonlinearities are distinct from Barkhausen noise. The ac
source has very low levels of noise and distortion, and the
analog-to-digital sampling is synchronized well with the ac drive.

\begin{figure}[!tb]
\begin{center}
\includegraphics[width=0.4\textwidth,clip]{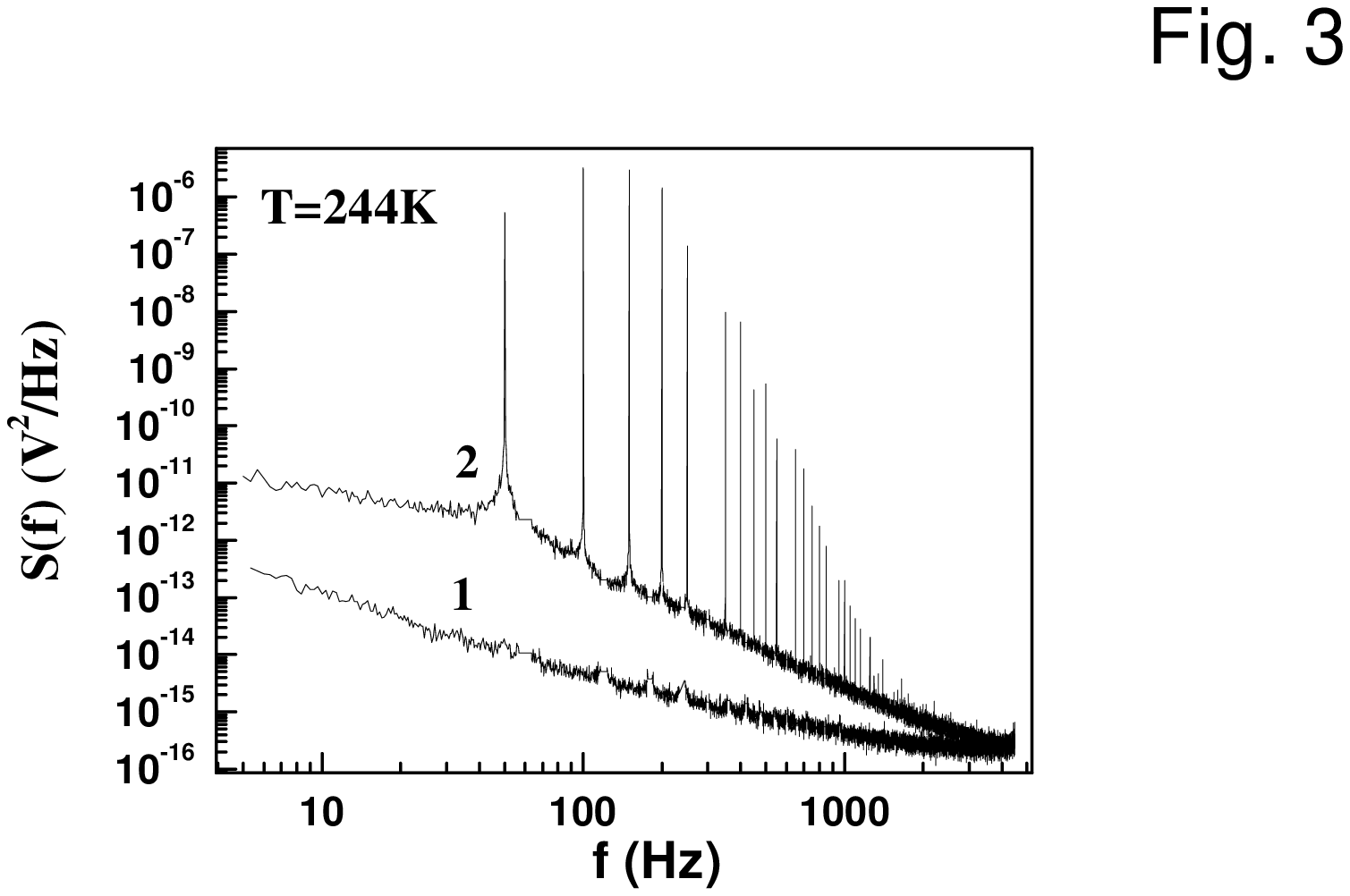}
\end{center}
\caption{
$S(f)$ with $E_D = 130$~V/cm at $f_D = 50$~Hz at 244~K (spectrum 2) is
shown above a background (spectrum 1) consisting of equilibrium noise
from the dissipative sample plus amplifier noise, mainly from current
noise from the Stanford 550 preamp. 60~Hz and its harmonics have been
edited out.}
\label{fig3}
\end{figure}

Fig.~\ref{fig3} shows a typical $S(f)$ with, roughly as expected, a
mixture of periodic and broad aperiodic noise. The wings around $f_D$
(and to a lesser extent around low harmonics) were not anticipated, but
these can arise from small drifts in the bridge cancellation as the
sample ages. The size of the polarization events, calculated below from
$S(f)$, was far too small for them to be discerned individually.

\begin{figure}[!tb]
\begin{center}
\includegraphics[width=0.4\textwidth,clip]{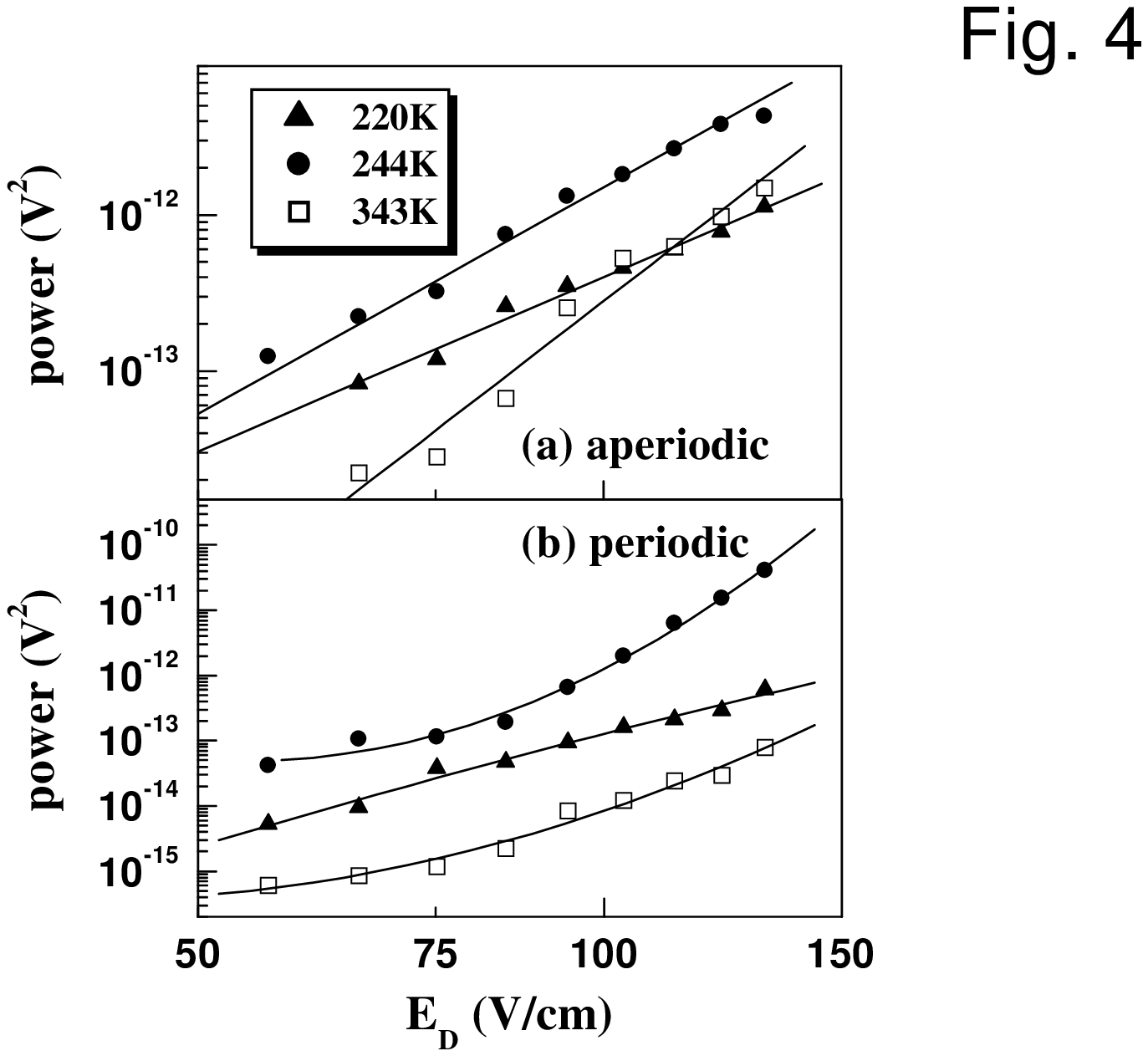}
\end{center}
\caption{
Noise power (above the relevant backgrounds) in the octave from 512
to 1024~Hz as a function of $E_D$ for (a) aperiodic and (b) periodic
components.}
\label{fig4}
\end{figure}

Since the spectral form is close to the independent-step expectation, we
can extract a typical step size $p$, although we doubt that the
polarization changes literally consist of simple discrete steps. As
shown in Fig.~\ref{fig4}, both periodic and aperiodic components of the
noise grow nonlinearly with $E_D$ over the range explored. Thus the $p$
calculated at some $E_D$ gives only a typical dipole step size under the
particular drive conditions. The form of the nonlinearity suggests a
broad distribution of $p$'s, with many small $p$'s showing up in the
noise only as $E_D$ is increased enough to make $f_C > f_D$ for that
$p$. The form of the aperiodic spectrum indicates a similar
distribution.

Although we are not yet confident in extracting absolute step sizes from
this new technique (mainly due to the difficulty of removing all
artifacts from the lower harmonics, which contain the most power), we
can use Eq.~\ref{eq:TotBkhNoise} to calculate a lower-bound estimate for
a typical $p$ by integrating the broad aperiodic spectrum only. (We
assume this $p$ reflects about how big the cooperative changes are
regardless of whether they are simple discrete steps.) At $T=250$~K,
just above the relaxor regime, at the largest $E_D$ used, the result is
about $p = 2 \times 10^{-22}$~C~cm, which apparently would grow at
higher $E_D$. Under the same conditions, the envelope of the harmonic
noise has a $1/f^2$ tail extending from about 1~kHz to 4~kHz, the top of
our measured range, indicating that there are some discrete dipole
switches as large as $2 \times 10^{-21}$~C~cm. An order-of-magnitude
estimate of the moment of a polar nanodomain at 250~K, based on the
range of the static polar correlations
\cite{Vakhrushev:96-You:00-Yoshida:98} and the saturation polarization
\cite{Bokov:60}, would be about $3 \times 10^{-23}$~C~cm. Thus it seems
that even in the \emph{paraelectric} regime there are some dynamically
coherent units larger than single polar nanodomains.

\begin{figure}[!tb]
\begin{center}
\includegraphics[width=0.4\textwidth,clip]{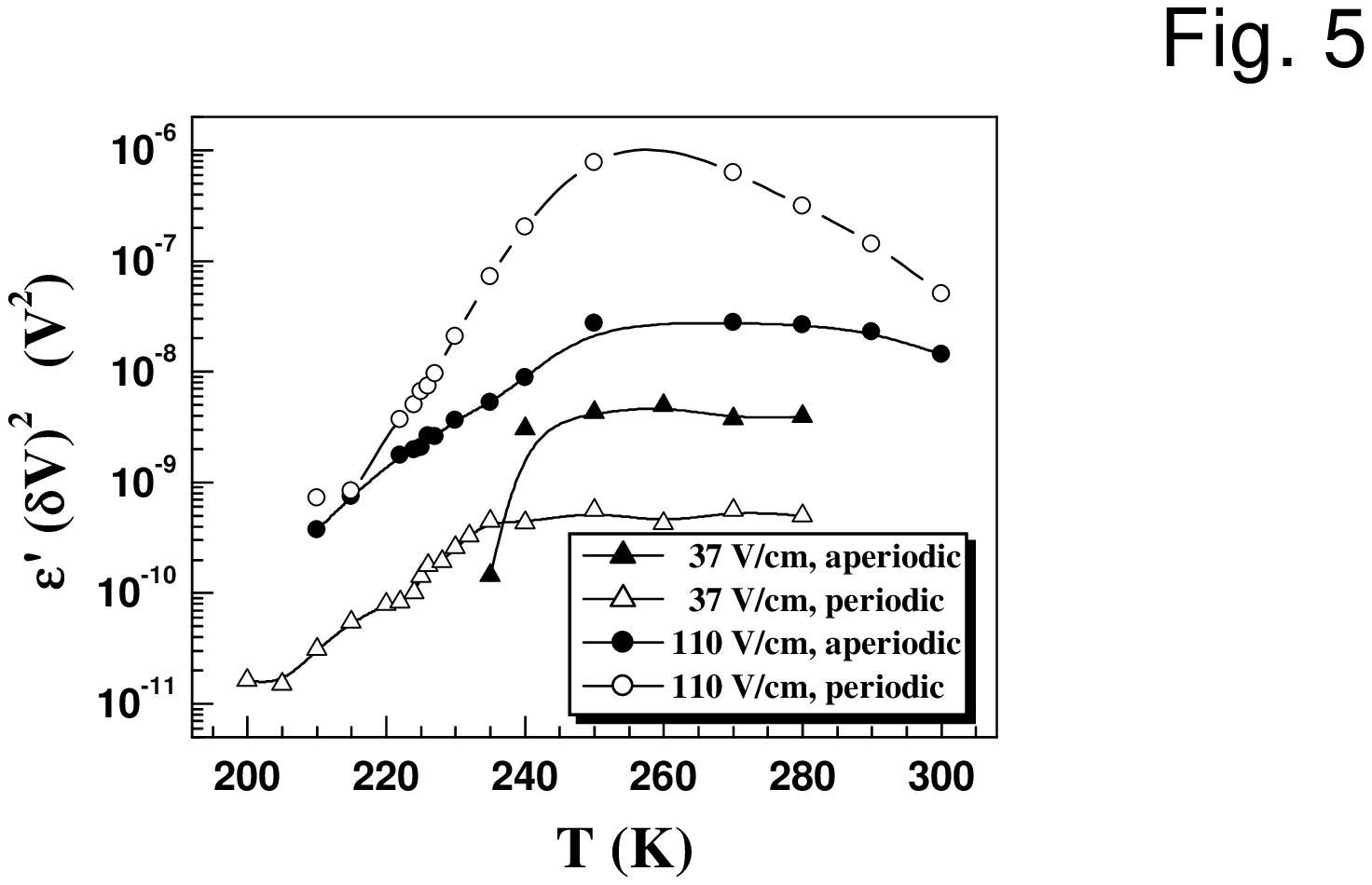}
\end{center}
\caption{
The magnitudes of periodic and aperiodic spectral components (above the
relevant backgrounds) in the range 512 to 1024~Hz, weighted by
$\varepsilon '(T)$, are shown vs. $T$ at two different $E_D$.}
\label{fig5}
\end{figure}

Despite the uncertainties in the absolute calibration, we can obtain a
good indication of how typical $p$'s depend on $T$, a key issue for
relaxor models. For independent steps, the quantity $\varepsilon ' \
(\delta V)^2$ is proportional to the actual $p$'s under particular
experimental conditions. Fig.~\ref{fig5} shows how some periodic and
aperiodic components of $\varepsilon ' \ (\delta V)^2$ depend on $T$ at
fixed $E_D$. We pick spectral components in an octave around $f = 15
f_D$, because in this range the signal-to-background is still good, we
expect few artifacts in the harmonics, and the $T$-dependence of $p$
shows up in $S(f)$ both via the overall magnitude and via the
$T$-dependence of $f_C$. (Other frequency components show the same sort
of behavior, just less clearly.)

As seen in Fig.~\ref{fig5}, as the relaxor regime is approached
from high $T$, the spectral components of $\varepsilon ' \  (\delta
V)^2$ either rise slowly, as one would guess from the growth of
medium-range static correlations \cite{Vakhrushev:96-You:00-Yoshida:98},
or show little change. However, below 235~K to 250~K, both periodic and
aperiodic components drop sharply, in sharp contrast to the continued
growth of static correlations. The factor $\varepsilon '(T)$ accounts for
less than half of this drop.

The width of the envelope of the harmonic components is also
$T$-dependent, being largest (for $E_D = 110$~V/cm) at 250~K. In
contrast to the long $1/f^2$ tail at 250~K, below 250~K the harmonics
fell sharply into the anharmonic background, e.g.\ becoming undetectable
above about 900~Hz at 210~K, confirming that the largest dipole moment
steps occur near 250~K.

In summary, our noise results indicate abrupt shrinkage of the typical
net dipole moment of the dynamically coherent changes (at fixed driving
field) as the relaxor regime is entered. We do not want to burden this
experimental result with much theory, but some initial comments should
be helpful.

We know of no proposed model in which the coherent dipole moment
differences between the \emph{equilibrium} states at different fields would
shrink sharply in the relaxor regime (i.e., relaxors are not believed to
be frustrated antiferroelectrics). Rather, the shrinkage of the steps
found on a fixed frequency scale should be related to the
Vogel-Fulcher kinetics \cite{Levstik:98,Vugmeister:00}, i.e., to the rapid
growth of barrier heights as
the relaxor regime is approached. Although our results do not dictate a
model for the relaxor transition in PMN, we can rule out some models.

A model of the dynamics with a fixed set of relaxation modes
with a fixed broad unimodal distribution of activation barriers
(presumably monotonically increasing in $p$) \cite{Tagantsev:94}, would
predict that the characteristic $p$ is a weakly increasing function of
$T$ at all $T$, and hence is inconsistent with our results. For simple
growth of independent polar nanodomains, the dipoles grow along with the
anisotropy barriers, so the typical dipole moment for which the field
(plus thermal activation) allows switches at rates of order $f_C$ would
continue to grow. For formation of coherently flipping clusters, the
typical moment also grows with the number of nanodomains (probably as a
square root). Although the moment to barrier ratio would be lower for
larger clusters, the moment of the largest clusters with rapid enough
driven kinetics would approximately saturate. Unless there were a
strange distribution of cluster sizes, so that as the typical size grew,
the mean of those remaining below some fixed cutoff would shrink, such
models would not reproduce our results.

Thus the Barkhausen results strongly suggest that the rapid growth of
barriers is not just the result of nanodomain growth or of collective
nanodomain clusters forming. Instead, it seems that each nanodomain's
contribution to the barriers grows much more rapidly than its
contribution to the dipole moment, as in a recent phenomenological model
\cite{Vugmeister:00}, suggesting that less-polar degrees of freedom are
freezing along with the polar nanodomain orientations. Another
indication of some such effects is that the low-temperature linear heat
capacity specifically associated with the relaxor \cite{Hegenbarth:95}
gives a dimensionless entropy of about $10^3$ per nanodomain.

Models of relaxor freezing invoking strong coupling of nanodomain
polarization to multiple slow local degrees of freedom (particularly
charge transfers) and to soft TO phonon modes \cite{Gehring:00} have
been proposed to account for clusters with multiple metastable
polarization states having a roughly uniform distribution in solid angle
\cite{Vikhnin:00FE}. Such models would allow complex aging
\cite{Colla(mbw):aging} even in small clusters of nanodomains. Whether
or not such models prove successful, the Barkhausen data provide a new
qualitative constraint on models for the relaxor transition, which has
thus far provided a target for somewhat under-constrained theory.

\section*{Acknowledgement}
This work was funded by NSF DMR 99-81869.

\end{document}